# Design of auction-based approach for market clearing in peer-to-peer market platform

*Mohsen Khorasany\*, Yateendra Mishra\*, Gerard Ledwich\**

*\* School of Electrical Engineering and Computer Science, Queensland University of Technology, Brisbane, Australia
Emails: m.khorasany@qut.edu.au, yateendra.mishra@qut.edu.au, g.ledwich@qut.edu.au*

## Abstract

This paper designs a market platform for Peer-to-Peer (P2P) energy trading in Transactive Energy (TE) systems, where prosumers and consumers actively participate in the market as seller or buyer to trade energy. An auction-based approach is used for market clearing in the proposed platform and a review of different types of auction is performed. The appropriate auction approach for market clearing in the proposed platform is designed. The proposed auction mechanism is implemented in three steps namely determination, allocation and payment. This paper identifies important P2P market clearing performance indices, which are used to compare and contrast the designed auction with different types of auction mechanisms. Comparative studies demonstrate the efficacy of the proposed auction mechanism for market clearing in the P2P platform.

## 1 Introduction

Today`s grid is characterized by increasing levels of distributed energy resources (DERs), demand response programs, and energy efficiency initiatives. With the increasing penetration of DERs, the traditional energy consumers have become prosumers, who can both produce and consume energy [1]. Increasing level of DERs confronts the grid with significant consequences, which introduces new challenges for distribution system operators [2]. Power distribution system needs a new framework to facilitate the use of DERs by enabling them to join traditional providers in production, buying and selling electricity [3]. Transactive Energy (TE) is such a novel framework and according to the GridWise Architecture Council (GWAC) is defined as "a system of economic and control mechanisms that allows the dynamic balance of supply and demand across the entire electrical infrastructure using value as a key operational parameter". TE systems can improve efficiency and reliability of the grid and help system operators to manage the increasing complexity of the grid. Implementing TE requires a clear mental model that is applicable and actionable. One area of concern for discussion of TE is policy and market design [4]. In the market design in TE systems, policies should be designed in a way that maximizes customer engagement via two-way interaction and accommodates the scale of DERs that will enter the system in the future. Design of this new market motivates prosumers and consumers to use electricity generated by renewable energy resources. A novel energy trading mechanism among prosumers, which can be used in TE systems is called Peer-to-Peer (P2P) energy trading which contributes to the balance of energy [5] and reduces congestions on transmission and distribution lines [6]. The P2P approach provides localized energy trading which facilitates system operation with a large penetration of DERs as inter-connected nodes in a network. In this context, a P2P market platform is presented in this paper, which enables direct energy trading among prosumers and consumers.

Design of localized energy trading platform has been investigated in some recent works. The overview of P2P communication technologies and different interoperability issues for the smart grids market platform has been reported in [7, 8]. A four-layered architecture for the smart grid is presented in [9], where the details of the market platform are not addressed. The decentralized P2P framework has been proposed in [10] and [11], where only renewable energy producers' perspective is considered. A self-sustainable community of prosumers is proposed by authors in [12], where a load aggregated on behalf of prosumers participate in the market. Although incentivize all DERs owners to participate in the market, these studies overlook the importance of satisfying both consumers and prosumers.

An important step in developing a P2P electricity market is the design of market clearing method. In the P2P market, energy allocation and electricity price should be determined in a way that incentivizes both prosumers and consumers to participate in the market. The auction-based approach can be used as market clearing mechanism, where it can be applied to any market with different numbers of sellers and buyers. Design of auction for different purposes has been addressed in some recent works. Authors in [13] propose an auction-based approach for sharing storage capacity among the residential community and shared facilities controller. Knapsack approximation algorithm is used in [14] as market clearing mechanism for a single seller and multiple buyers. A Vickrey-Clarke-Groves (VCG) mechanism is proposed in [15] to maximize the social welfare in demand side management. Social welfare of plug-in hybrid electric vehicles is maximized in a localized P2P electricity trading using an iterative double auction in [16]. Authors in [17] propose two variants of VCG mechanism to charge plug-in hybrid electric vehicles in residential distribution networks. A novel two-phase auction format for on-line scheduling of demand side aggregation is presented in [18].

In this paper, an auction-based market clearing approach is presented, which is unique in its ability to enable sellers and buyers to trade energy through a platform based on a set of rules and satisfying both sellers and buyers at the same time. A review of the most recognized types of auctions is performed to investigate important features of auctions and then, an appropriate auction mechanism is designed for market clearing to satisfy both prosumers and consumers in the market. A set of indices is defined to compare the proposed auction mechanism with other types of auction. The main contributions of this paper include the following aspects:
- Propose a market platform for P2P energy trading in TE environment
- Design auction approach for market clearing in the proposed platform
- Compare performance of the designed auction with different types of auction for market clearing in the proposed platform

This paper is structured as follows: The P2P market framework is presented in Section 2. The required features and rules of the auction for market clearing in this system are cleared in Section 3, followed by a review of different types of auction. The proposed auction for market clearing in the P2P market is designed in Section 4. Finally, the numerical analysis is provided in Section 5 and some concluding remarks are drawn in Section 6.

## 2 P2P market platform

This paper proposes a four-layered framework based on the GWAC TE framework for P2P energy trading among sellers and buyers in distribution network [3]. Sellers are prosumers with excess energy having their own objective to maximize their revenue in the market, whereas buyers can be either prosumers or consumers. Different layers of the proposed framework are decision making layer, business layer, cyber layer and component layer. In decision making layer, sellers offer the amount of energy and their reservation price, whereas buyers bid for their demanded energy. Network constraints are monitored in the data centre of the P2P market, which is connected to Distribution System Operator (DSO). An unregulated environment, where energy related services are freely produced, traded, sold and consumed among several market players through the smart market is provided in the business layer. The cyber layer contains smart meter, data centre and all communication infrastructures to enable the proposed function of the transactions in the business layer. The component layer is the physical distribution of all participating components in the distribution grid.

In this framework, market clearing is performed in the business layer, where an auction-based approach is considered to clear the market in a way that incentivizes market players to use this framework. The proposed P2P market is an hourly auction, where sellers offer their surplus energy to the market and buyers bid for their demand. In designing market for the TE framework, the following assumptions are considered:
- Buyers/sellers know the market price and Feed-in Tariff (FIT) and submit their bids/offers truthfully based on these quantities.
- Buyers can buy their required amount of energy from multiple sellers and consequently, sellers can sell their energy to several buyers.
- The total network is divided into several blocks and prosumers who can transfer energy to each other using existed network are members of the same block and participate in the same auction.
- Buyers and sellers in the same block can use the existed network to transfer energy without paying a transfer fee.
- Each seller/buyer submits one offer/bid and any other offer/bid would be considered as from separate seller/buyer for auction purposes.

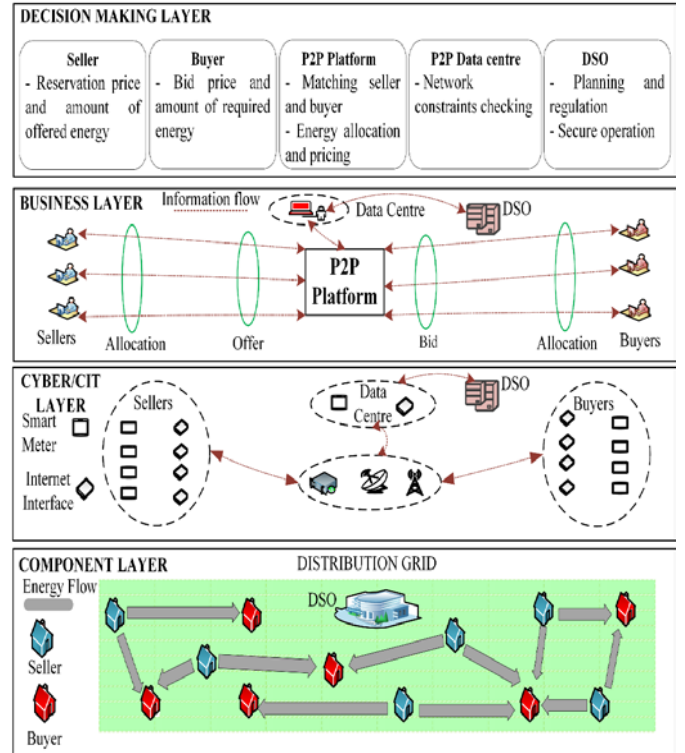

Figure 1: Energy trading framework in distribution network

Consider a market with a set of $\mathcal{N} = \{1, ..., N\}$ sellers and $\mathcal{M} = \{1, ..., M\}$ buyers who participate in the auction. The offer of seller $i$ indicates the amount of energy ($x_i$) and reservation price for this energy ($r_i$), while bid of buyer $j$ represents the demanded energy ($q_j$) and the offered price for this energy ($b_j$). The objective in the P2P market is to maximize the revenue of sellers and cost saving of buyers at the same time. Therefore, the total revenue and cost saving of all players (TRC) is defined as (1):

$$TRC = \sum_{i=1}^{N} R_i + \sum_{j=1}^{M} C_j \qquad (1)$$

where $i$ and $j$ are indices of seller and buyer respectively. The revenue of seller $i$ ($R_i$) and cost saving of buyer $j$ ($C_j$) can be defined as (2) and (3):

$$R_i = \sum_j (\lambda_{i,j} - r_i) x_{i,j} \qquad (2)$$
$$C_j = \sum_i (b_j - \rho_{j,i}) q_{j,i} \qquad (3)$$

where $x_{i,j}$ and $\lambda_{i,j}$ are the amount of sold energy by seller $i$ to buyer $j$ and its corresponding price respectively, whereas $q_{j,i}$ and $\rho_{j,i}$ represents the amount and price of bought energy by buyer $j$ from seller $i$. In the proposed P2P market, the allocation and price of energy is indicated using auction approach. The market clearing for each time slot "$t$" is performed in the previous time slot "$t-1$" as follow:

*Step 1*: Collecting offers/bids from sellers/buyers and market price and FIT from local generator.
*Step 2*: Conducting auction and indicating allocation and price of energy
*Step 3*: Informing sellers/buyers for the final allocation and cleared market price.

An appropriate auction mechanism is required in Step 2 to enable competitive P2P energy trading which is discussed in the next section.

## 3 Auction approach for market clearing

### 3.1 Auction properties and rules

The auction is defined as "a well-specified negotiation mechanism mediated by an intermediary that can be considered as an automated set of rules" [19]. Auctions have several properties and can be categorized based on their features. The most important features of auction are:

- Open or sealed-bid: In an open auction, bidders know about other participants' bids, but in the sealed-bid auction, the bid of bidders is not publicly known.
- One-sided or two-sided: In one-sided auction, only buyers submit bids, but in the double auction, both buyers and sellers submit bids.
- Single or multi-unit auction: In single unit auction buyers bids for one unit, whereas in multi-unit auction buyers bids for multiple units.
- Strong balanced budget (SBB) or Weak balanced budget (WBB): In the SBB auction the auctioneer should not lose or gain money, while in the WBB auction the auctioneer can gain money, but should not lose money.

Appropriate auction mechanism for use in the P2P market is a sealed-bid, two-sided and multi-unit auction. This auction is SBB as there is no real auctioneer and energy trading is performed using a platform. Each auction has specific rules and buyers and sellers should participate in the auction based on these rules. The auction for the proposed P2P framework is implemented in three steps namely *determination* where the number of sellers and buyers who can participate in the auction is identified; *Allocation* where energy share for seller and buyers are resolved; *Payment* is where the auction prices for buyers and sellers for traded energy are settled.

### 3.2 Review of different types of auction

In this paper, different types of auction and their properties are compared to find the appropriate auction for P2P market clearing. The most recognized types of auction in the electricity market are one-sided and only buyers bid in the auction such as uniform price, pay as bid Vickrey, and Generalized second price auction. However, there are other types of auction where both buyers and sellers participate (Double or two-sided auction) and the determination step indicates that only buyers with bids higher than reservation price of sellers can participate in the auction, but the number of participant sellers ($L$) and buyers ($K$) is different in various auction types. The most recognized double auctions are double auction with average mechanism, VCG mechanism, trade reduction mechanism, and McAfee`s mechanism.

Allocation method in all auctions is based on "greedy" algorithm and energy from a seller with lower offer (reservation price) ($r_i$) is allocated to buyers with higher bid ($b_j$). Number of participant sellers/buyers in the auction is the same in all double auction except double auction with trade reduction and McAfee mechanism, where only $L$-$1$ sellers and $K$-$1$ buyers trade energy. The most important difference between various types of double auction is in the payment rule, in which in the double auction with average mechanism, all buyers should pay the average of reservation price of sellers and bid of buyers. In Trade reduction mechanism the auction price is $r_L$ for sellers and $b_K$ for buyers, whereas, in McAfee mechanism, the payment is based on the average of $r_{L+1}$ and $b_{K+1}$ [20]. The only auction which has the required properties for P2P market clearing is a double auction with average mechanism. In the next section, a new auction mechanism for market clearing is designed and then the performance of this mechanism is compared with all of the aforementioned mechanisms.

## 4 Auction design for market clearing in P2P market

### 4.1 Determination rule

The determination rule for the proposed auction is the same as the aforementioned auction types, where after collecting all bids and offers from buyers and sellers, the determination rule identifies the number of sellers and buyers selected to participate in the auction as below:
- Sellers' declare their reservation price $r_i$ and these prices are arranged in increasing order as $r_1 < r_2 < \cdots < r_N$; whereas buyers' bids are arranged in decreasing order as $b_1 > b_2 > \cdots > b_M$.
- Aggregated supply and demand curve is generated.
- Let $k$ be the largest index such that $b_K \geq r_L$ (the "breakeven point"). The first $L$ sellers sell the energy to the first $K$ buyers.

## 4.2 Allocation rule

A greedy algorithm is used for allocation of energy from sellers to buyers, where energy from the seller with lowest reservation price is allocated to the buyer with the highest bid. However, the procedure of matching buyers and sellers depends upon the method of players' participation in the auction. A seller/buyer can participate in the market in two ways, e.g. fractional or non-fractional participation where seller/buyer can win any fraction of its offer/bid or total offer/bid after market clearing respectively. The allocation of energy can be performed from sellers perspective or buyers perspective, where $x_i^*$ and $q_j^*$ indicates the sold/bought energy by seller/buyer in the market respectively. The allocation from the buyer/seller perspective is based on fractional knapsack problem, where buyers/sellers can decide to whether participate in the market fractionally or non-fractionally.

The major difference between these two algorithms is the priority in the allocation of energy, in which in the first scenario the priority is to allocate energy to buyers to satisfy their demand, but in the second scenario, priority is to allocate energy from sellers to reach their offered energy. The allocation algorithm from buyers' perspective is shown in table 1. The final allocation of energy is different depends on the total offered energy by sellers ($\sum_{i=1}^{L} x_i$) and total demanded energy by buyers ($\sum_{j=1}^{K} q_j$). If $\sum_{i=1}^{L} x_i = \sum_{j=1}^{K} q_j$, the final allocation is the same for two algorithms. And if $\sum_{i=1}^{L} x_i < \sum_{j=1}^{K} q_j$, the allocation is performed from buyers` perspective, to indicate the method of participation of the last buyer; otherwise allocation will be performed from sellers` perspective to enable the last seller to decide on the method of his participation.

## 4.3 Payment rule

The next step after energy allocation is identifying the price of traded energy. The designed market should be balanced for both sellers and buyers to incentivize them to participate in the P2P market. Therefore, the total revenue of sellers should be equal to the total cost saving of buyers as in (4).

$$\sum_{i=1}^{L} R_i = \sum_{j=1}^{K} C_j \quad (4)$$

Also, as discussed in Section 3, the appropriate auction for the P2P market is SBB, which means that the total money paid by buyers goes to sellers i.e. for each individual $i$ and $j$

$$\lambda_{i,j} = \rho_{j,i} \quad (5)$$

*Theorem 1-* The auction has a balanced performance for both sellers and buyers if the price of energy is indicated by (6):

$$\lambda_{i,j} = \frac{r_i + b_j}{2} \quad (6)$$

*Proof-* By substituting (2) and (3) in (4) and expanding, it can be written as (7)

$$\sum_j (\lambda_{1,j} - r_1) x_{1,j} + \cdots + \sum_j (\lambda_{L,j} - r_L) x_{L,j}$$
$$= \sum_i (b_1 - \rho_{1,i}) q_{1,i} + \cdots + \sum_i (b_K - \rho_{K,i}) q_{K,i} \quad (7)$$

where it can be expanded again as (8)

$$x_{1,1}(\lambda_{1,1} - r_1) + \cdots + x_{L,K}(\lambda_{L,K} - r_L) = $$
$$q_{1,1}(b_1 - \rho_{1,1}) + \cdots + q_{K,L}(b_K - \rho_{K,L}) \quad (8)$$

---

1: Arrange sellers in ascending order and buyers in descending order
2:     **for** *j* starts from *1* to *K* **do**
3:         **for** *i* starts from *1* to *L* **do**
4:             **if** $q_j \leq x_i$ **then**
5:                 $q_j^* = q_j$ and update $x_i$
6:             **else if** $q_j > x_i$ AND $j \neq K$ **then**
7:                 Add $x_i$ till $q_j \leq \sum x_i$
8:                 $q_j^* = q_j$ and update $x_i$
9:             **else if** $q_j > x_i$ AND $j = K$ **then**
10:                 **if** $q_K \leq \sum_{i=1}^{L} x_i$ **then**
11:                     $q_K^* = q_K$ and update $x_i$
12:                 **else if** $q_K > \sum_{i=1}^{L} x_i$ **then**
13:                     $q_K^* = 0$ if buyer is non-fractional
14:                     $q_K^* = \sum_{i=1}^{L} x_i$ if buyer is fractional
15:                 **end if**
16:             **end if**
17:         **end for**
18:     **end for**
19: The $(x_i^*, q_j^*)$ is achieved.

Table 1: Allocation algorithm from buyers' perspective

The sold energy by seller *i* to buyer *j* is equal to the bought energy by buyer *j* from seller *i*. Therefore, for each individual *i* and *j*,

$$x_{i,j} = q_{j,i} \quad (9)$$

Substituting (5) and (9) in (8) yields that the price of traded energy should be equal to average of $r_i$ and $b_j$ for each individual *i* and as shown in (6).

*Theorem 2-* In the proposed market, if the used auction for market clearing is SBB, the total revenue of sellers and cost saving of buyers is independent of the price of traded energy.

*Proof-* TRC can be written in the expanded format as (10).

$$TRC = \sum_{i=1}^{L} R_i + \sum_{j=1}^{K} C_j$$
$$= \sum_j (\lambda_{1,j} - r_1) x_{1,j} + \cdots + \sum_i (b_K - \rho_{K,i}) q_{K,i}$$
$$= x_{1,1}(\lambda_{1,1} - r_1) + \cdots + q_{K,L}(b_K - \rho_{K,L}) \quad (10)$$

Substituting (7) yields

$$TRC = x_{1,1}(\lambda_{1,1} - r_1 + b_1 - \rho_{1,1}) + \cdots + x_{L,K}(\lambda_{L,K} - r_L + b_K - \rho_{K,L}) \quad (11)$$

Again by substituting (8), TRC can be rewritten as (12), which shows its independence to the price of traded energy.

$$TRC = \sum_{i=1}^{L} \sum_{j=1}^{K} x_{i,j} (b_j - r_i) \quad (12)$$

## 4.4 P2P market performance indices

In addition to TRC, another important index in designing an auction mechanism is the satisfaction of market players at the end of each time slot. Seller and Buyer Satisfaction Index (SSI and BSI) can be defined as (13) and (14) respectively.

$$SSI_i = \frac{\sum_{j=1}^{K} x_{i,j} \lambda_{i,j}}{x_i r_i} \quad (13)$$

$$BSI_j = \frac{q_j b_j}{\sum_{i=1}^{L} q_{j,i} \rho_{i,j}} \quad (14)$$

SSI/BSI shows the proportion of the real income/cost of seller/buyer to the expected income/cost. The higher value of SSI and BSI indicates higher satisfaction of seller and buyer respectively and SSI and BSI lower than 1, show dissatisfaction of them. Market Tendency Index (MTI), which shows the skewness of the market is determined by averaging of satisfaction of all market players as defined in (15). If the value of MTI is greater than 1, the market is more beneficial for buyers and it is skewed toward sellers if MTI is less than 1.

$$MTI = \frac{\sum_{j=1}^{K}\left(\frac{BSI_j \sum_{i=1}^{L} q_{j,i}}{K}\right)}{\sum_{i=1}^{L}\left(\frac{SSI_i \sum_{j=1}^{K} x_{i,j}}{L}\right)} \quad (15)$$

## 5 Numerical analysis

This section provides the numerical analysis to evaluate the performance of the proposed P2P market and the designed auction mechanism for market clearing. One auction round is considered for a block with 16 players (eight buyers and eight sellers). In this market, based on determination rule, the first five sellers and five buyers can trade energy for the next time slot in the auction, i.e. $K = L = 5$ and the last three sellers and buyers lose the auction. Input data including sellers' submitted offers and buyers' submitted bids are tabulated in Table 2. Based on the discussion in the previous section, since $\sum_{i=1}^{5} x_i = \sum_{j=1}^{5} q_j = 700$, the final allocation is the same in both allocation algorithms (sellers` perspective and buyers` perspective).

The allocation of energy for buyers and sellers and its corresponding price in the designed auction are given in Table 3. In this table, the amount of traded energy between each seller and buyer is specified using the allocation algorithm, where energy from sellers with lower reservation price is allocated to the buyers with higher bid and the price is obtained by (6). Auction prices in different auction mechanisms are presented in Table 4 to have a base scheme for comparison with the proposed mechanism. These prices are specified using offers/bids of seller/buyers and based on the payment mechanism in each auction. It is noteworthy that in this case study, auction price for McAfee mechanism is the same as trade reduction mechanism and consequently the final results would be the same for both mechanisms. Therefore, the following discussion and results for trade reduction mechanism are valid for McAfee mechanism.

A comparative study is performed for different auctions and TRC, MTI, revenue of sellers and cost saving of buyers for these auctions are given in Figure 2. TRC for double auction with VCG mechanism is the highest, however, this mechanism is not SBB, and the amount of received money by sellers is more than paid money by buyers. The lowest TRC is for double auction with trade reduction mechanism (also McAfee mechanism), where based on auction rules, seller *L*, and buyer *K* cannot trade energy and total traded energy decreases. This mechanism is not SBB too because the money paid by buyers is higher than received money by sellers. As it said in the previous section since in P2P market there is no real agent as moderator, the used auction should be SBB and all monetary transfers should be done between buyers and sellers. TRC for other payment mechanisms is the same, as proved in *Theorem 2*.

Total revenue of sellers in pay as bid and the generalized second price is higher in comparison to other payment mechanisms. In pay as bid mechanism, the total cost saving of buyers is zero and this mechanism is beneficial for sellers only. The highest total cost saving of buyers occurs in Vickrey mechanism. The total cost saving of buyers is equal to revenue of sellers in the proposed mechanism which shows the balanced performance of this mechanism. Figure 2 also shows MTI of the different payment mechanism. The Pay as bid mechanism has the lowest MTI which means this market is more beneficial for sellers and this mechanism is the best one from the sellers perspective. MTI in the Vickrey auction has the highest value (greater than 1) that shows the market is more beneficial for buyers. Double auction with average mechanism has the nearest MTI to 1, while MTI in the proposed mechanism is equal to 1, which means this auction has a balanced performance for both sellers and buyers and is appropriate for use in the P2P market platform.

| Seller/Buyer No. | 1 | 2 | 3 | 4 | 5 | 6 | 7 | 8 |
|---|---|---|---|---|---|---|---|---|
| Sellers` offers $r_i$ (¢/kWh) | 10.0 | 10.5 | 11.0 | 12.0 | 12.1 | 12.5 | 13.0 | 13.2 |
| $x_i$(Wh) | 200 | 150 | 100 | 150 | 100 | 100 | 150 | 100 |
| Buyers` bids $b_j$ (¢/kWh) | 14.0 | 13.5 | 13.0 | 12.5 | 12.2 | 12.0 | 11.5 | 11.0 |
| $q_j$(Wh) | 150 | 150 | 200 | 100 | 100 | 100 | 100 | 100 |

Table 2: Input data; offers/bids by sellers/buyers

| Amount of traded energy between seller *i* and buyer *j* and its corresponding price $(x_{i,j} = q_{j,i})(Wh)/(\lambda_{i,j} = \rho_{j,i})$(¢/kWh) | | | | | | | | | | |
|---|---|---|---|---|---|---|---|---|---|---|
| | S1 | | S2 | | S3 | | S4 | | S5 | |
| B1 | 150 | 12.00 | 50 | 11.75 | 0 | 11.50 | 0 | 11.25 | 0 | 11.10 |
| B2 | 0 | 12.25 | 100 | 12.00 | 50 | 11.75 | 0 | 11.50 | 0 | 11.35 |
| B3 | 0 | 12.50 | 0 | 12.25 | 100 | 12.00 | 0 | 11.75 | 0 | 11.60 |
| B4 | 0 | 13.00 | 0 | 12.75 | 50 | 12.50 | 100 | 12.25 | 0 | 12.10 |
| B5 | 0 | 13.05 | 0 | 12.80 | 0 | 12.55 | 0 | 12.30 | 100 | 12.15 |

Table 3: Allocation/price of energy in the proposed auction

| Payment Mechanism | Auction Price for sellers (¢/kWh) | | | | | Auction price for buyers (¢/kWh) | | | | |
|---|---|---|---|---|---|---|---|---|---|---|
| | $\lambda_{1,j}$ | $\lambda_{2,j}$ | $\lambda_{3,j}$ | $\lambda_{4,j}$ | $\lambda_{5,j}$ | $\rho_{1,i}$ | $\rho_{2,i}$ | $\rho_{3,i}$ | $\rho_{4,i}$ | $\rho_{5,i}$ |
| Uniform price | 12.20 | 12.20 | 12.20 | 12.20 | 12.20 | 12.20 | 12.20 | 12.20 | 12.20 | 12.20 |
| Vickrey | 12.00 | 12.00 | 12.00 | 12.00 | 12.00 | 12.00 | 12.00 | 12.00 | 12.00 | 12.00 |
| Average mechanism | 12.08 | 12.08 | 12.08 | 12.08 | 12.08 | 12.08 | 12.08 | 12.08 | 12.08 | 12.08 |
| VCG Mechanism | 12.20 | 12.20 | 12.20 | 12.20 | 12.20 | 12.10 | 12.10 | 12.10 | 12.10 | 12.10 |
| Trade reduction | 12.10 | 12.10 | 12.10 | 12.10 | - | 12.20 | 12.20 | 12.20 | 12.20 | - |
| Payment Mechanism | Auction Price for sellers (¢/kWh) | | | | | Auction price for buyers (¢/kWh) | | | | |
| | $\lambda_{i,1}$ | $\lambda_{i,2}$ | $\lambda_{i,3}$ | $\lambda_{i,4}$ | $\lambda_{i,5}$ | $\rho_{1,i}$ | $\rho_{2,i}$ | $\rho_{3,i}$ | $\rho_{4,i}$ | $\rho_{5,i}$ |
| Pay as bid | 14.00 | 13.50 | 13.00 | 12.50 | 12.20 | 14.00 | 13.50 | 13.00 | 12.50 | 12.20 |
| Generalized Second Price | 13.50 | 13.00 | 12.50 | 12.20 | 12.00 | 13.50 | 13.00 | 12.50 | 12.20 | 12.00 |

Table 4: Auction price for different auction types

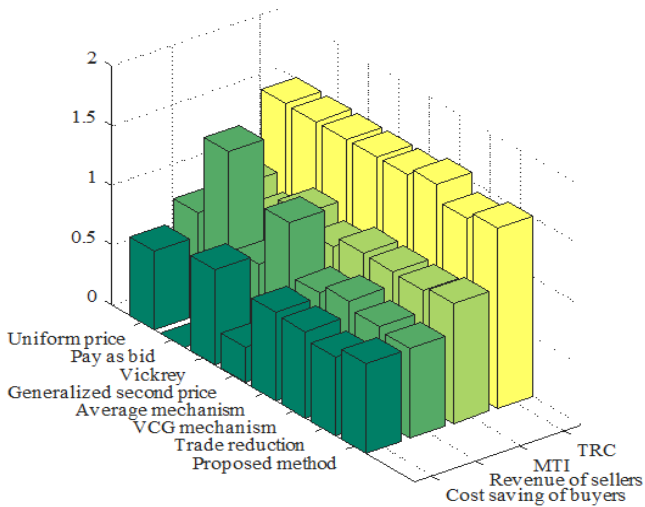

Figure 2: Graphical comparison of TRC, Sellers` revenue and buyers` cost saving

## 6 Conclusions

This paper proposes a novel market platform for P2P energy trading in TE framework. In this platform, prosumers and consumers participate in the market as seller or buyer. An auction-based approach for market clearing in the proposed platform is presented to enable sellers and buyers to trade energy through a platform based on a set of rules instead of requiring any real third party. A novel auction mechanism is designed for satisfying both prosumers and consumers in the market at the same time and required rules of auction in the P2P market are determined. Numerical indices are defined to compare the performance of the proposed auction with different types of auction mechanism for the P2P market. Numerical results verify that the proposed mechanism satisfies all required features for P2P market clearing and has a balanced performance for both sellers and buyers. In our upcoming work, we are going to use the proposed auction based approach in conjunction with an optimization problem to take into account the network constraints and proposing a more accurate market clearing for P2P energy trading.